\documentclass[prl,aps,showpacs,twocolumn]{revtex4}
\usepackage{epsfig}
\usepackage{bm}

\begin{document}

\title{Logarithmically modified  scaling of temperature structure functions in thermal convection}

\author{\small A. Bershadskii$^{1,2}$, K.R. Sreenivasan$^{1}$ and J.J. Niemela$^1$}
\affiliation{\small {\it $^1$International Center for Theoretical
Physics, Strada Costiera 11, I-34100 Trieste, Italy}\\
{\it $^2$ICAR, P.O. Box 31155, Jerusalem 91000, Israel}}

\begin{abstract}
Using experimental data on thermal convection, obtained at a
Rayleigh number of 1.5 $\times 10^{11}$, it is shown that the
temperature structure functions $\langle \Delta T_{r}^p \rangle$,
where $\Delta T_r$ is the absolute value of the temperature
increment over a distance $r$, can be well represented in an
intermediate range of scales by $r^{\zeta_p} \varphi (r)^{p}$,
where the $\zeta_p$ are the scaling exponents appropriate to the
passive scalar problem in hydrodynamic turbulence and the function
$\varphi (r) = 1-a(\ln r/r_h)^2$. Measurements are made in the
midplane of the apparatus near the sidewall, but outside the
boundary layer.
\end{abstract}


\maketitle

A deep similarity between statistical properties of turbulent
passive scalar fluctuations \cite{sa} and those of temperature
fluctuations in the turbulent thermal convection \cite{kad} has
been discovered recently from numerical simulations (see, for
instance, refs.\ \cite{ching1},\cite{biskamp} and the literature
cited therein). In the present note, we will use experimental data
and a brief theoretical reasoning to go further in this same
direction.

We consider turbulent convection in a confined container of
circular crosssection and 50 cm diameter. The aspect ratio
(diameter/height) is unity. The sidewalls are insulated and the
bottom wall is maintained at a constant temperature, which is
slightly higher than that of the top wall. The working fluid is
cryogenic helium gas. By controlling the temperature difference
between the bottom and top walls, as well as the thermodynamic
operating point on the phase plane of the gas, the Rayleigh number
($Ra$) of the flow could be varied between $10^7$ and $10^{15}$.
We measure the temperature fluctuations $T(t)$ at various Rayleigh
numbers towards the upper end of this range, in which the
convective motion is turbulent. The results used in this paper
correspond mainly to $Ra=1.5\times 10^{11}$. Time traces of
fluctuations are obtained at a distance of 4.4 cm from the
sidewall on the center plane of the apparatus. This position is
outside of the boundary layer region for the Rayleigh numbers
considered here. More details of the experimental conditions and
measurement procedure can be found in ref.\ \cite{NS1}.

For the structure functions $\langle \Delta T_{\tau}^p \rangle$,
where the absolute values of temperature increment
$\Delta T |T(t+\tau)-T(t)|$ is evaluated as a function of the time interval
$\tau$ and the angle brackets indicate time averaging, we do not
observe any clear scaling (see fig.\ 1). However, the normalized
structure functions show unambiguous scaling as
\begin{figure}\vspace{-0.5cm}
\centering \epsfig{width=.45\textwidth,file=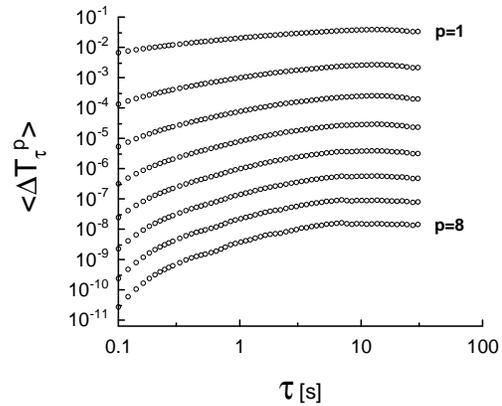}
\vspace{-4.5cm} \caption{Temperature structure functions $\langle
\Delta T_{\tau}^p \rangle$ against $\tau$ (in log-log
coordinates). Here and for other figures, $Ra = 1.5 \times
10^{11}$. }
\end{figure}
$$
S_p (\tau) =\frac{\langle \Delta T_{\tau}^p \rangle}{\langle
\Delta T_{\tau}^2 \rangle^{p/2}}  \sim \tau^{-z_p}   \eqno{(2)}
$$
(see fig.\ 2). The scaling exponents $z_p$, extracted as slopes of
the best straight-line fits shown in fig.\ 2 (the scaling range is
from $0.1$s to $10$s), indicate multiscaling. These exponents are
shown in fig.\ 3 as circles.

It should be noted that temperature structure functions in thermal
convection were investigated in ref.\ \cite{ching2} using the
experimental data described in \cite{sano}. In particular, the
data used in \cite{ching2} were obtained at the {\it center} of
the convection cell. While the structure functions $\langle \Delta
T_{r}^p \rangle$ themselves showed no scaling properties, two
scaling ranges were observed for the normalized structure
functions $S_p(\tau)$. In the present paper, we study the data
obtained near sidewall (see above), where sufficiently strong
convection wind is present. For this case, we observe only one
scaling interval for $S_p(\tau)$.

Since, for a passive scalar $\Theta$ advected and diffused by
hydrodynamic turbulence, the scaling of structure functions
follows (in both experiments and direct numerical simulations) the
relation
$$
\langle \Delta \Theta_{\tau}^p \rangle \sim \tau^{\zeta_p},
     \eqno{(3)}
$$
the corresponding exponents of the {\it normalized} structure
functions can be readily calculated from the definition as
$$
z_p = \zeta_p -\frac{p}{2} \zeta_2.  \eqno{(4)}
$$
We also show in fig.\ 3 the values of $z_p$, calculated using
equation (4), for the passive scalar measured in the atmosphere
\cite{schmidt} (inverted triangles) and from a direct numerical
simulation of three-dimensional homogeneous isotropic turbulence
\cite{wg} (upright triangles). They are virtually
indistinguishable from each other and from the present exponents
extracted for thermal convection from fig.\ 2.
\begin{figure}
\centering \epsfig{width=.45\textwidth,file=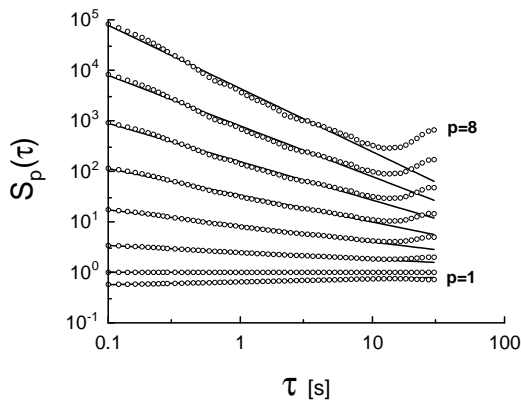}
\vspace{-4.5cm} \caption{Normalized temperature structure
functions $\langle \Delta T_{\tau}^p \rangle /\langle \Delta
T_{\tau}^2 \rangle^{p/2}$ against $\tau$ (in log-log coordinates).
Straight lines (the best fits) are drawn to indicate the
multiscaling (2).}
\end{figure}

Noting this good correspondence between $z_p$ for thermal
convection and for the passive scalar, and further that there is
no scaling of unnormalized structure functions in convection, we
suggest that the structure functions $\langle \Delta T_{\tau}^p
\rangle$ in convection should have the general {\it non-scaling}
form
$$
\langle \Delta T_{\tau}^p \rangle \sim \tau^{\zeta_p} \varphi
(\tau)^{p},  \eqno{(5)}
$$
where the exponents $\zeta_p$ are taken from the passive scalar
problem. This suggestion is completely consistent with the
experimentally observed multiscaling properties (2)-(4) (fig.\ 3).
One can find $\varphi (\tau)$ explicitly from the data by dividing
$\langle \Delta T^p_{\tau} \rangle $ by $\tau^{\zeta_p}$ and
taking the $p$-th root of the resulting function. Figure 4 shows
the outcome of these calculations for different values of $p$ (the
calculated function $\varphi (\tau)$ is normalized by its
maximum). In the semi-log scales used in the Fig.\ 4, the $\varphi
(\tau)$ function has an essentially parabolic form given by
$$
\varphi (\tau) = [1-a (\ln \tau /\tau_h )^2] \eqno{(6)}
$$
where $a$ and $\tau_h$ are constants. The scatter of the data in
fig.\ 4 needs a brief comment. The range of $\tau$ shown in the
plot covers more than two decades. Although the tails (especially
for large orders $p$) do not collapse perfectly, most of the data
(in the range $-2 < \ln \tau < 2.5$, i.e. about two decades of
$\tau$) possess a scatter of less than $6\%$.

The case of particular interest, corresponding to the second
order, is shown in the inset to fig.\ 4; the arguments due to
Corrsin and Obukhov \cite{my}) yield a specific value for the
second order ($\zeta_2 = 2/3$ without intermittency corrections).
The solid parabola (the best fit) corresponds to (6) with $a\simeq
0.04$ and $\tau_h \simeq 0.64s$. A slightly different choice for
$\zeta_2$ to account for the intermittency effects shows no
essential difference.

\begin{figure}
\centering \epsfig{width=.45\textwidth,file=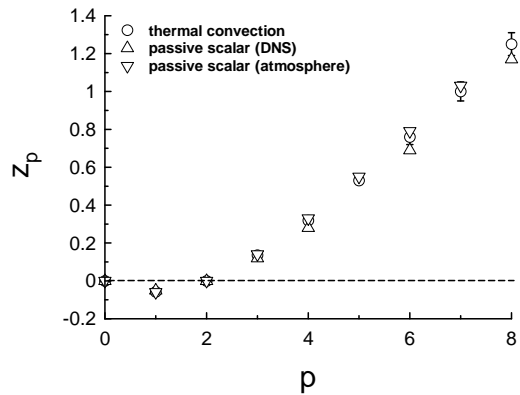}
\vspace{-4.5cm} \caption{Scaling exponents $z_p$, for the
normalized structure functions $S_p (\tau)$, extracted from fig.\
2 (circles). The scaling exponents $z_p$ calculated using equation
(4) for the passive scalar in the atmosphere \cite{schmidt}
(inverted triangles) and in the direct numerical simulations of
mixing in three-dimensional homogeneous isotropic turbulence
\cite{wg} (upright triangles). }
\end{figure}

It may be useful to make the following observation. In general,
for the atmosphere, the structure function data are interpreted
with respect to spatial separation through the use of Taylor's
hypothesis. Because of the presence of the mean wind in the
present experiment, one may use Taylor's hypothesis \cite{my} with
equal facility and interpret temperature increments over time
intervals as increments over equivalent spatial distances
traversed by the mean wind. This hypothesis is neither critical
nor necessary, though we shall use it later to be in conformity
with standard practice. If the Taylor's hypothesis is applied to
temperature fluctuations in convection, one would replace the time
separation $\tau$ by an equivalent spatial separation $r$, and we
would have
$$
\langle \Delta T_{r}^p \rangle \sim r^{\zeta_p}\cdot [1-a(\ln
r/r_h)^2]^p. \eqno{(7)}
$$

Local scale invariance characteristic of structure functions with
scaling properties can often be interpreted, if only loosely, in
terms of a conformal invariance, which may be understood as an
extension of the Kolmogorov-Obukhov similarity hypothesis. There
is evidence that this conformal invariance is broken for the
velocity field in three-dimensional hydrodynamic turbulence
\cite{lp}. Equation (7) suggests the Corrsin-Obukhov scaling
possesses a symmetry-breaking property (cf also \cite{mull}).

To understand the ``correction" of the passive scalar scaling in
(5) it is useful to consider the equations of motion in the
Boussinesq approximation:
$$
Pr^{-1} (\frac{\partial}{\partial t}+ {\bf v}\cdot \nabla) {\bf v}
= -\nabla p + T \hat{z}+ Ra^{-1/2} \nabla^2 {\bf v} \eqno{(8)}
$$
$$
(\frac{\partial}{\partial t}+ {\bf v}\cdot \nabla) T = Ra^{-1/2}
\nabla^2 T
  \eqno{(9)}
$$
$$
\nabla \cdot {\bf v} =0.  \eqno{(10)}
$$
Here ${\bf v}$ and $p$ are the velocity and pressure fields,
$\hat{z}$ is a unit vector in the vertical direction and $Pr$ is
the Prandtl number. We expect that the anisotropy related to the
thermal origin of velocity fluctuations is diminished in the
inertial range of scales with $Ra \rightarrow \infty$ and so may
regard, for large $Ra$, the temperature fluctuations in the spirit
of a ``perturbation" over the passive scalar problem.

\begin{figure}
\centering \epsfig{width=.45\textwidth,file=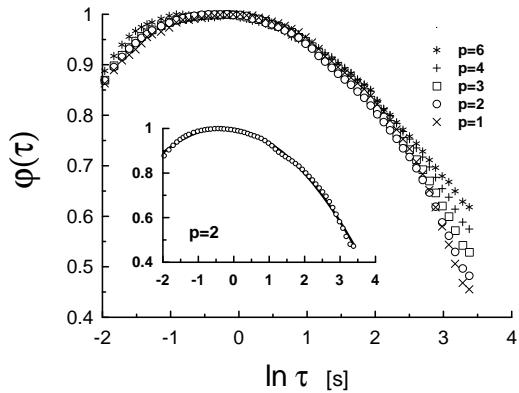}
\vspace{-4.5cm} \caption{The function $\varphi (\tau)$ in semi-log
scales. The solid parabola indicates equation (6).}
\end{figure}

\begin{figure}
\centering \epsfig{width=.45\textwidth,file=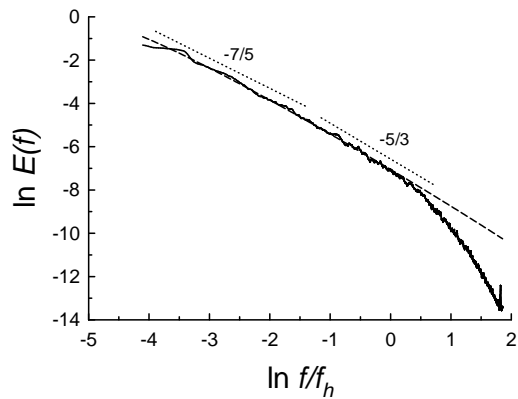}
\vspace{-4.5cm}\caption{The spectral density of the temperature
fluctuation (solid curve) against normalized frequency (in log-log
coordinates). The dashed curve corresponds to the logarithmically
corrected Corrsin-Obukhov spectrum (12) and dotted straight lines
indicate the Bolgiano-like form $E(f) \sim f^{-7/5}$ and the
Corrsin-Obukhov form $E(f) \sim f^{-5/3}$. }
\end{figure}

Finally, we wish to address briefly the power spectrum of
temperature fluctuations. There is no simple way of relating the
form (7) for the second-order structure function to a similarly
neat form for the spectrum. However, considering that the
correction to scaling presumably has its origin in a perturbation
theory, we will use the same form as (7) for the spectrum as well,
and replace $r$ by $k^{-1}$ (where $k$ is wave number) to write
$$
E(k) \sim k^{-5/3} [1-a(\ln k/k_h)^2],   \eqno{(11)}
$$
or, for the frequency spectrum,
$$
E(f) \sim f^{-5/3} [1-a(\ln f/f_h)^2],  \eqno{(12)}
$$
where $f_h=\tau_h^{-1}$ and the power ``$-5/3$" is the
Corrsin-Obukhov scaling for the passive scalar---corresponding to
$\zeta_2 =2/3$ noted above \cite{my}. Figure 5 shows the
experimentally observed spectrum of the temperature fluctuations
({\it solid} curve). The logarithmically corrected Corrsin-Obukhov
form (12) is shown in this figure as the {\it dashed curve}. (The
fitting range, clearly seen in fig.\ 5, is a bit smaller than for
the normalized structure functions, where a part of "dissipation"
range is also covered by the fitting.) We also indicate by {\it
dotted} lines the Corrsin-Obukhov form $E(f) \sim f^{-5/3}$ and
the Bolgiano-like form $E(f) \sim f^{-7/5}$, mentioned previously
in the literature in relation to the turbulent thermal convection
(refs.\ \cite{sano},\cite{y}-\cite{nssd}). However, the
Bolgiano-like fit of the spectrum has no support in the
second-order structure function in our case.

In conclusion, our proposal is that the temperature fluctuations
in the convection problem can be treated as a ``perturbation" over
the passive scalar problem, and that the ``correction" is of the
non-scaling form shown in (7). Empirical evidence in favor of this
proposal is presented. Developing this suggestion into a
full-fledged theory is beyond the scope of the present article.

\acknowledgments

We thank L. Biven, A. Praskovsky, V. Steinberg, and V. Yakhot for
useful discussions and help in calculations, T. Watanabe and T.
Gotoh for providing their paper \cite{wg} before publication, and
the referees for valuable comments.

\end{document}